\documentstyle[aps,prl,amstex,twocolumn,psfig]{revtex} 

\begin{document} 
 
\draft 
\title{Faraday waves on a viscoelastic liquid} 
\author{C.~Wagner$^1$, H.~W.~M\"uller$^{2,3}$, and K.~Knorr$^1$} 
\address{$^1$ Institut  f\"ur Technische Physik , Universit\"at 
des Saarlandes Postfach 151150, D-66041 Saarbr\"ucken, Germany\\
$^2$ Institut f\"ur Theoretische Physik, Universit\"at des Saarlandes
 Postfach 151150, D-66041 Saarbr\"ucken, Germany\\
$^3$ Max Planck Institut f\"ur Polymerphysik, Ackermannweg 10,
D-55128 Mainz } 
\maketitle 
 
\begin{abstract} 
We investigate Faraday waves on a viscoelastic liquid. 
Onset measurements and a nonlinear phase diagram for the selected patterns 
are presented. By virtue of the elasticity of the material a surface resonance
synchronous to the external drive competes with the usual 
subharmonic Faraday instability. Close to the 
bicriticality the nonlinear wave interaction gives rise
to a variety of novel surface states: Localised patches of hexagons,
hexagonal superlattices, coexistence of hexagons and lines. 
Theoretical stability calculations
and qualitative resonance arguments support the experimental observations.
\end{abstract} 
% 
% 47.50.+d Non-Newtonian fluid flows
% 47.54.+r  Pattern selection; pattern formation
% 47.20.-k : Hydrodynamic stability 
% 47.10.+g : General theory (of fluid dynamics) 
% 47.20.Ky : Nonlinearity (including bifurcation theory) 
% 47.20.Ma Interfacial instability 
% 47.20.Gv Viscous instability 
% 47.20.-k Hydrodynamic stability 
% 47.15.Cb Laminar boundary layers 

%**************************************************************** 
%*                                                              * 
%*      begin of text                                           * 
%*                                                              * 
%**************************************************************** 
\narrowtext  
The generation of standing waves on the surface of a vertically vibrated
Newtonian fluid (Faraday waves) is one of the classical hydrodynamic pattern forming
instabilities \cite{faraday31}. 
Theoretical and experimental investigations during the last
decade have substantially improved our understanding of the underlying 
processes \cite{miles90}.  Recently  pattern formation in viscoelastic
  fluids came into
 the focus  of nonlinear science, too. The memory  of  viscoelastic fluids introduces an  
additional time scale, which gives rise to 
a number of interesting phenomena \cite{vest69,brand86,pleiner88,larson90,groisman96}.  
It is amazing that already Faraday \cite{faraday31} tried to compare 
Newtonian fluids with viscoelastic ones: 
{\it
        "The difference between oil and white of egg
        is remarkable; (...) 
        the crispated state may be a useful and even important indication of
        the internal constitution of different fluids."
        }
In Newtonian fluids the prevailing surface mode of the Faraday experiment 
is usually subharmonic (S),
i.e. the waves oscillate with twice the period of the external excitation.
Nevertheless, for very thin layers 
 the surface oscillation may become harmonic (H), i.e. synchronous to the 
 external drive \cite{cerda97,muller97}. However, 
the necessary vibration conditions are rather extreme, rendering this
parameter region difficult to explore.
 
 On the other hand, viscoelastic Maxwell fluids have recently been predicted to
exhibit the harmonic Faraday resonance  \cite{muller98a}, if
the external drive frequency compares to the elastic relaxation 
 time of the fluid. This statement is based on the elasticity of the 
material and independent of the filling level. Thus selecting a polymeric 
liquid with an
 appropriate elastic time scale will open a new way to 
a systematic investigation of the harmonic Faraday instability.
 The present paper deals with such an experiment. The predicted
 harmonic resonance competes with the usual 
 subharmonic instability. Close to the bicriticality an interesting new
 surface dynamics occurs, unknown in Newtonian fluids. 

Our working fluid is a $1.5\%$ 
(by weight) solution  of  
PAA (polyacrylamide-co-acrylic acid of molecular 
weight $5 \times 10^6$, Aldrich 18,127-7) 
in  a $60:40$ glycerol-water solvent. 
The tabulated values for density, surface tension and dynamic viscosity 
 of the {\em solvent} at $20^\circ C$
are $\rho=1,154 kg/m^3$, $\sigma = 0.069 N/m$ and $\eta=11mPas$. For the 
polymeric solution the complex dynamic viscosity $\eta^\star(f)=\eta'(f)-i\eta''(f)$  
 has been measured at room temperature ($\simeq 22^\circ C$) 
with a rotational viscometer (Rheometrics Fluids). Data acquisition for
frequencies above $f=16Hz$ was not possible. However,
within the range $1Hz<f<16Hz$ 
 the data could well be fitted by the following  
 power laws: $\eta'=0.082 \times f^{-0.594} Pas$
and $\eta''=0.122 \times f^{-0.549} Pas$, where $f$ is measured in $Hz$. 
These relations have been used to extrapolate the viscosity data
into the frequency range of our experiment ($30Hz<f<100Hz$). 

The  experimental setup consist of a black container built out of anodised aluminium  
 with an  
inner diameter of 290 mm, filled to a height of 3 mm with the working fluid.   
 By means of an electromagnetic shaker (V617 Gearing \& Watson) the container 
is vibrated vertically with an acceleration $a \, \cos{\Omega t}$, 
where $\Omega=2 \pi f$. The container is covered and sealed by a glass plate to 
prevent 
evaporation and pollution. No degradation or drift of the experimental results
could be detected within a period of four weeks. 
Due to the heat production of the vibrator
 we were not able to work at room temperature, where the 
rheometric viscosity data have been taken.   
 A heat wire filament regulates the temperature at $30 ^\circ C \pm 0.1 $,
measured by a  PTC-resistor imbeded in 
the bottom of the vessel. Waveform generation as well as data acquisition are performed by a 
D/A-board (National Instruments, AT-MIO 16 E2) in a Pentium PC. The vibration
signal measured  
by a piezoelectric device (Bruel \& Kjear 4393) was checked to be sinusoidal with
contributions from higher harmonics not more than $2\%$.
A ring of 150 LED's with twice the container diameter is mounted on the vertical
axis of the setup and illuminates the surface. The light reflected into
the centre of the ring is 
recorded by a full frame CCD camera (Hitachi KPF-1). Its electronic shutter
 allows to synchronise the exposure with the external drive signal an thus to 
discriminate the H and S surface response.
The phase diagram of the selected patterns is obtained by quasi-statically
ramping the drive amplitude from just
 below the  
onset $\epsilon=\frac{a}{a_c}-1=-5\%$ 

 %*******************************************************************

\begin{figure}
\psfig{file=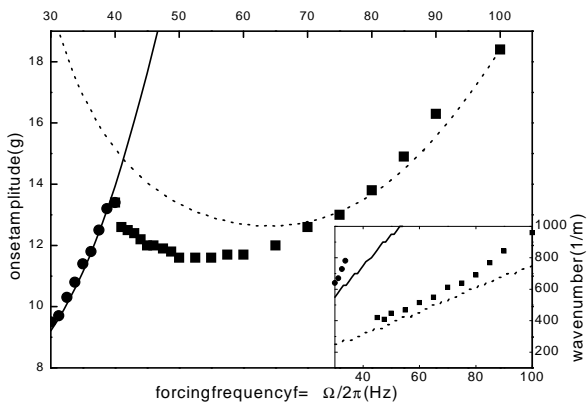,width=0.95\columnwidth,angle=0} 
  
\caption[] 
{ 
The acceleration amplitude $a_c$ and critical wave numbers $k_H$ and
$k_S$  for the
onset of the harmonic (circles) and subharmonic (squares) Faraday instability. 
Experimental data are plotted by symbols, lines denote the theoretical 
computation on the basis of the rheometric viscosity data $\eta^\star(f)$. 
} 
\label{fig1} 
\end{figure} 
%********************************************************************

up to $\epsilon=10\%$  while keeping the frequency constant.
 During each scan the amplitude is increased by $1\%$-steps
 and held constant for $300$ seconds in between.
Then, the surface of the fluid is photographed
and the time dependence  of the surface oscillation (whether $S$ or $H$)
is determined. Similar downwards-amplitude scans are 
 performed to detect a possible hysteresis. 
  
Fig.~\ref{fig1} shows the threshold amplitude ${a_c}$ for the critical
wave numbers of the 
Faraday instability. In the frequency domain 
above $f\simeq 60Hz$  
one observes the usual subharmonic 
Faraday resonance with a surface pattern of lines. Since this
 behavior is similar
to high viscosity Newtonian liquids \cite{edwards94,chen97}, we denote this
region as the Newtonian regime.  
The present paper, however,
is focused on  the frequency range below $40 Hz$, where the 
 surface responds harmonically. The predominance of the harmonic 
Faraday resonance has recently been predicted for viscoelastic 
Maxwell fluids \cite{muller98a}. On using the rheometric 
viscosity data  for our working fluid 
we have computed the stability threshold (lines in Fig.~\ref{fig1}) by the    
method of Kumar and
Tuckerman \cite{kumar94}. Favourable agreement is achieved except for frequencies close
to the bicritical intersection point. There are two possible sources for discrepancies: (i) 
There is a $7-9^\circ C$  difference between the temperature of
our experimental setup and that of the rheometer during the viscosity measurement.
  (ii) Errors due to the extrapolation
of the viscometric data into the frequency range of our experiment.

\begin{figure} 
\psfig{file=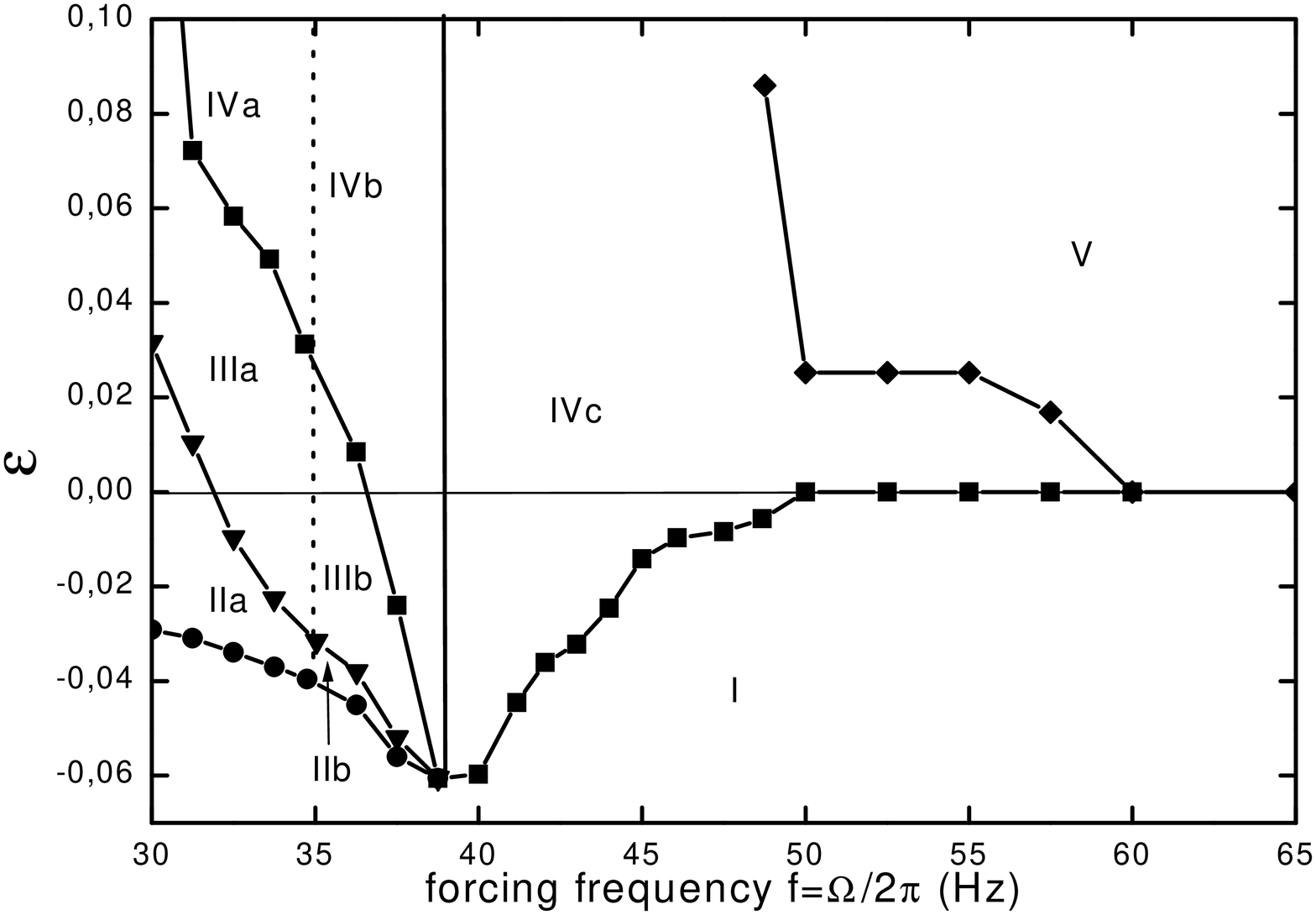,width=0.95\columnwidth,angle=0} 
\caption[] 
{ 
Phase diagram of the observed 
nonlinear patterns. The symbols mark experimental data points. 
The abscissa, $\varepsilon=0$,  indicates the linear threshold 
shown in Fig.~\ref{fig1}. 
The lowest solid line denotes the
saddle point of the hysteresis. Region $I$: flat surface;  
$IIa$: harmonic hexagons covering the whole surface; $IIIa$: harmonic-subharmonic
hexagonal superlattice (see text) extending over the whole surface; $IIb,IIIb$: as before but
{\em localised patches} surrounded by the flat surface;  $IVa,IVb,IVc$: chaotic 
dynamics of subharmonic 
lines competing with extended (a) or localised (b) hexagonal superlattices or the flat 
surface (c); $V$: stationary subharmonic lines extending over the whole surface
(Newtonian regime). 
} 
\label{fig2} 
\end{figure}

\begin{figure} 
\psfig{file=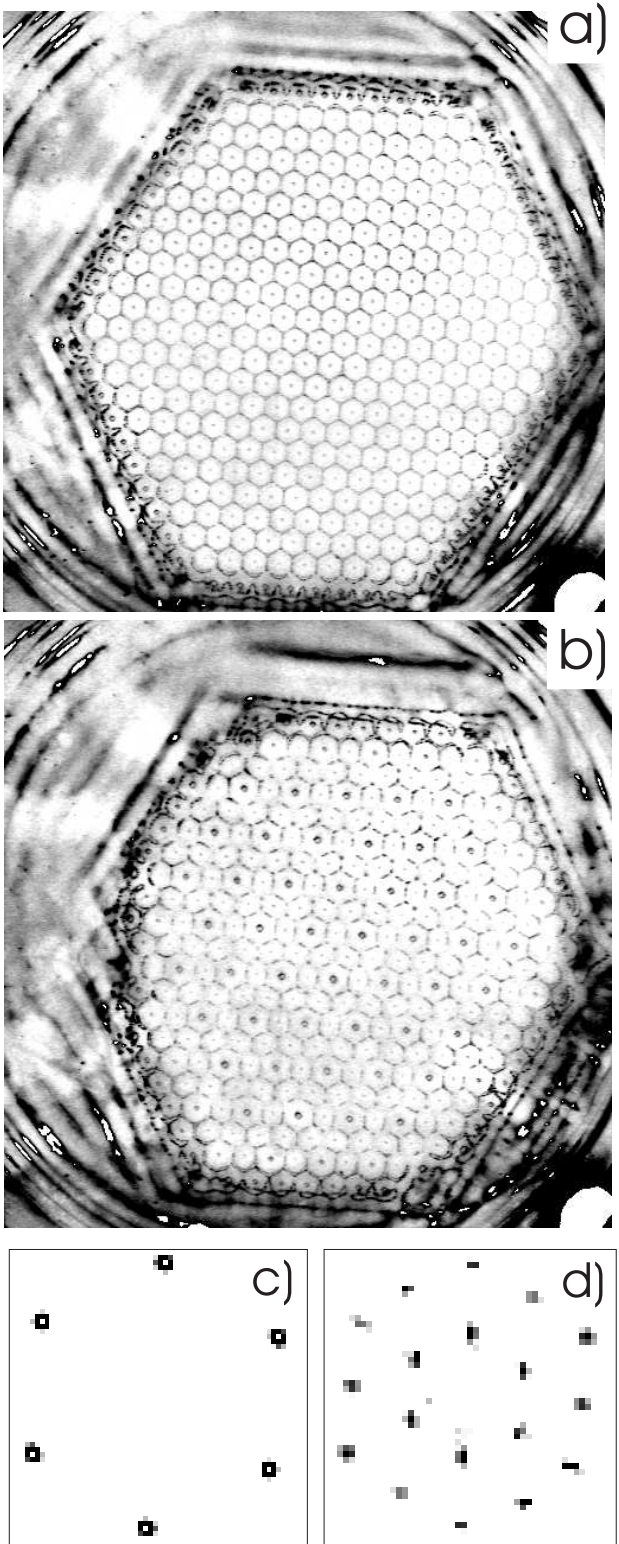,width=0.95\columnwidth}                             
\caption[] 
{ 
Localised stationary surface patterns of harmonic hexagons (a) and the 
harmonic-subharmonic 
hexagonal superlattice (b) as observed in region $IIb$ and $IIIb$ at $f=37 Hz$ and
$\varepsilon=-0.05$ and $-0.03$. The corners of the picture show 
the outer edge of the container. In (c) and (d) the respective 
spatial Fourier spectra are shown. Patterns in regions $IIa$ and $IIIa$ are similar
but they cover the whole surface. 
} 
\label{fig3} 
\end{figure}

Fig.~\ref{fig2} depicts the phase diagram of the selected patterns. The 
surface structure in region $IIa$ is a uniform pattern of 
hexagons covering the whole surface, while in $IIb$ we observe very 
stable localised patches of hexagons (see Fig.~\ref{fig3}a).
A slight onset hysteresis up to 
$6\%$ 
in $\varepsilon$ is visible.
The pattern selection processes for 

 %************************************************************************
%****************************************************************************

the backwards bifurcating hexagons in the 
harmonic region $II$ 
and the forwards bifurcating lines in the subharmonic Newtonian regime 
$V$ rely on the distinct 
respective time dependencies. This can be seen as follows: On taking the 
surface elevation $\eta$ as a representative order parameter one has
\begin{eqnarray}
\label{surf}
\eta({\bf r},t) =  \left [ \sum_{m} \left \{ \begin{array}{r@{\quad}l}
    H_m e^{i {\bf k}_{Hm} \cdot {\bf r}} \\ 
    S_m e^{i {\bf k}_{Sm} \cdot {\bf r}} \end{array} \right \} + c.c. \right ]
\times \\
\sum_{n=-\infty}^{+\infty}  \eta_n 
\left \{ \begin{array}{r@{\quad}l}  e^{i n \Omega t}  \\ 
                                           e^{i (n+ \frac{1}{2}) \Omega t} 
\end{array} \right \} \quad \mbox{for} \quad 
\left \{ \begin{array}{r@{\quad}l} H \\ S
\end{array} \right \}. \nonumber
\end{eqnarray}
Here ${\bf r}=(x,y)$ is the horizontal coordinate, 
the lateral wave vectors ${\bf k}_{Hm}$ and 
${\bf k}_{Sm}$ with $|{\bf k}_{Hm}|=k_H$ and 
$|{\bf k}_{Sm}|=k_S$ compose the spatial
pattern and the $\eta_n$ are the temporal Fourier coefficients
determined by the linear stability problem 
(components of the Floquet eigenvector). A similar ansatz holds for the 
velocity field ${\bf v}$.
On feeding $\eta$ and ${\bf v}$ into an arbitrary 
quadratic nonlinearity of the hydrodynamic equations results in a 
frequency spectrum of integral multiples of 
$\Omega$, no matter  whether S or H is considered. Thus 
quadratic nonlinearities are able to resonate  with {\em harmonic} 
linear eigenmodes, but not with {\em subharmonic} ones. Clearly, spatial
resonance must be granted as well. Thus
 any triple of harmonic modes 
$\{{\bf k}_{H1},{\bf k}_{H2},{\bf k}_{H3}\}$ with $|{\bf k}_{Hm}|=k_H$ and
${\bf k}_{H1}+{\bf k}_{H2}+{\bf k}_{H3}=0$ is in resonance. This generic 
3-wave vector 
coupling is well known
e.g. from Non-Boussinesq-Rayleigh-B\'{e}nard convection and 
enforces a saddle node

%*****************************************************************************
%*********************************************************************************

 bifurcation towards 
hexagonal patterns. The associated 
amplitude equations 
are of the form  (use cyclic permutation to get the other ones; 
$\kappa$ and  $\gamma$ are nonlinear coefficients)
\begin{equation}
\partial_t H_1=\varepsilon H_1 + \kappa H_2^\star H_3^\star -
\left \{ |H_1|^2+\gamma\left[|H_2|^2+|H_3|^2\right ] \right \}H_1.
\label{hexagons}
\end{equation}
Within the Newtonian region $V$ such a  resonant 
3-wave vector coupling is prohibited due to the missing  temporal resonance. 
Quadratic nonlinearities in the evolution equations for the 
$S_m$ are prohibited and the pattern selection mechanism is
controlled by the cubic nonlinearity \cite{muller93,chen97}.  
It is instructive to point out that the 
subharmonic {\em temporal} symmetry
$\eta(t+2\pi/\Omega)=-\eta(t)$ for Faraday waves is analogous to
the {\it spatial} up-down symmetry for Rayleigh-B\'{e}nard 
convection (known as the
Boussinesq symmetry): They both
prevent quadratic nonlinearities to appear in the
associated amplitude equations. 

If the drive amplitude is raised from $IIa,b$ to 
$IIIa,b$, a sharp secondary 
transition can be detected: Subharmonic frequency contributions suddenly 
appear  in the temporal Fourier spectrum of the surface elevation. 
 Simultaneously the associated spatial Fourier spectrum 
(Fig.~\ref{fig3}c,d) 
exhibits  a hexagonal superstructure with the same orientation but a wavelength 
twice as long as the primary one.  The
theoretical stability calculation indicates
that the {\em primary} subharmonic onset (dashed line in Fig.~\ref{fig1})
is situated far above the $II-III$ transition. Therefore,
the secondary superstructure has to result from a
nonlinear excitation process. The underlying mechanism is again a 3-wave vector
interaction: Invoking basic resonance arguments the subharmonic mode 
${\bf k}_{S1}=\frac{1}{2} {\bf k}_{H1}$ with
amplitude $S_1$ obeys the 
following amplitude equation (use again cyclic permutation to get the equations
for $S_2$ and $S_3$)
\begin{equation}
\label{eqn3}
\partial_t S_1=- \lambda  S_1 + \chi H_1 S_1^\star +  \mbox{higher order terms.}
\end{equation}
Here $\lambda>0$ reflects the linear damping and $\chi$ is a nonlinear
coupling coefficient. According to Eq.~(\ref{eqn3}) the secondary 
crossover from $II$ to $III$ occurs
when the  primary pattern amplitude
$|H_m|=H$ exceeds the threshold
$\lambda  / |\chi|$. 

The dashed lines in 
Fig.~\ref{fig2} separating the subregions of homogeneous ($a$ ) and
localised ($b$) hexagons are to be considered 
as approximate boundaries
rather than sharp transition lines.  On increasing
$f$, the localised patches decrease in size until they
dissappear at $f>39Hz$. However, on keeping $f$ fixed their size  
is rather robust and hardly depends on the forcing amplitude $a$. 
Note that these new localised patches of hexagons cannot be explained by 
a set of 
Ginsburg-Landau equations, which supplements Eqs.~(\ref{hexagons}) by diffusive
 spatial derivatives:  Within this familiar model isolated islands of 
hexagons do not {\em stably} exist over a finite control parameter interval.

 %*********************************************************************
   \begin{figure} 
   \psfig{file=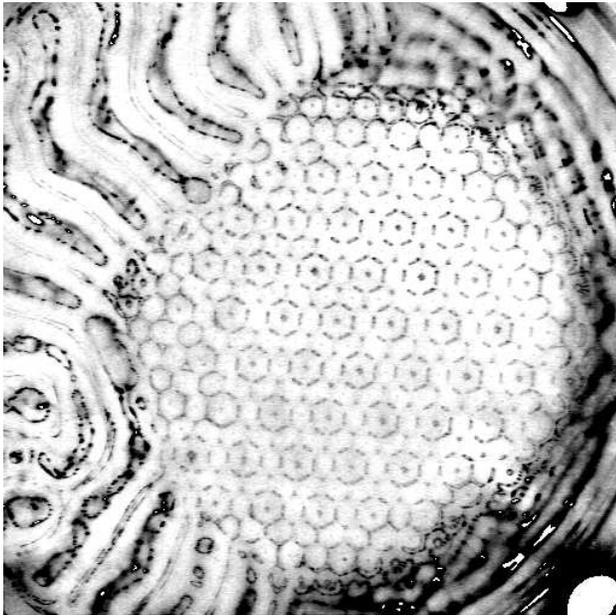,width=0.95\columnwidth} 
  % \parbox[t]{8 cm}
\caption[] 
{ 
Snapshot of a time dependent surface state in region $IVb$ at $f=37 Hz$ and $\varepsilon=2$. 
A stationary localised patch of hexagons is surrounded and competes with 
subharmonic lines moving in an erratic manner across the surface. 
} 
\label{fig4} 

\end{figure} 
%************************************************************************

On entering region $IVa$ or $IVb$ the patterns become chaotically time dependent.
 Patches of subharmonically 
oscillating lines originating in an erratic manner from 
the cell boundary $(a)$ or respectively from the
flat surface $(b)$ penetrate into the stationary 
hexagonal superlattice. Then they disappear again and the original structure 
 is recovered.
 This dynamics repeats on time scales of the order of seconds 
to minutes, leading 
to a temporary coexistence  of the stationary 
hexagonal superlattice with subharmonic lines (see Fig.~\ref{fig4}). 
Higher drive amplitudes lead to a fully chaotic surface pattern.

For drive frequencies in the intermediate region $Vc$  a similar
chaotic behavior of subharmonic lines can be observed, however directly 
as the {\em  primary} instability. Since
hexagons do not occur beyond $f\simeq39Hz$, the subharmonic lines compete 
with the flat surface state. 
 Nevertheless, the bifurcation still exhibits a small hysteresis, which continuously 
goes down as $f$ approaches $60 Hz$. 
%**************************************************************** 
%*                                                              * 
%*      Summary                                                 * 
%*                                                              * 
%**************************************************************** 

In summary,   
we have presented experimental results on Faraday waves in a viscoelastic
medium. Earlier theoretical predictions of the harmonic Faraday resonance
could be confirmed. 
Our measurements are the first systematic study of  
harmonic Faraday waves, which could not be performed yet for Newtonian fluids. 
The empiric onset data are found in good quantitative agreement with the theory. 
Furthermore the
appearance of uniform hexagonal patterns and their (secondary) transition 
towards a harmonic-subharmonic
superlattice is understood in terms of elementary $3-$wave vector resonances. 
 Besides spatially homogeneous wave patterns
we find very robust and sharply localised patches, which are 
surrounded by the flat surface or -- at higher drive amplitudes -- in complicated 
dynamical competition with
subharmonic lines. Their explanation requires a more
elaborate theory.
  
%**************************************************************** 
%*                                                              * 
%*      Acknowledgements                                        * 
%*                                                              * 
%**************************************************************** 
 
{\it Acknowledgements\/} ---  We thank H.~Rehage for the rheometric 
viscosity measurements and helpful comments. Furthermore we acknowledge
support by J.~Albers. This work is supported by the  Deutsche 
Forschungsgemeinschaft.

%**************************************************************** 
%*                                                              * 
%*      list of references                                      * 
%*                                                              * 
%**************************************************************** 


\begin{thebibliography}{99} 
%
 \bibitem{faraday31}M.~Faraday, Philos. Trans. R. Soc. London {\bf 52},  
319 (1831). 
%  
\bibitem{miles90} for a review see: J.~W.~Miles and D.~Henderson, Ann. Rev. Fluid Mech.  
{\bf 22}, 143 (1990); H.~W.~M\"{u}ller, R.~Friedrich, and D.~Papathanassiou, 
{\it Theoretical and experimental studies of the Faraday instability}, in:
{\it Lecture notes in Physics}, ed. by F.~Busse and S.~C.~M\"{u}ller,
Springer (1998). 
% 
\bibitem{vest69} 
C.~M.~Vest and V.~S.~Arpaci, J. Fluid Mech. {\bf 36}, 613 (1969);  
M.~Sokolov and R.~I.~Tanner, Phys. Fluids {\bf 15}, 534 (1972).  
% 
\bibitem{brand86} 
H.~R.~Brand and B.~J.~A.~Zielinska, Phys. Rev. Lett. {\bf 57}, 3167 (1986); 
B.~J.~A.~Zielinska, D.~Mukamel and V.~Steinberg,
Phys. Rev. E {\bf 33}, 1454 (1986). 
% 
\bibitem{pleiner88} 
H.~Pleiner, J.~L.~Harden and P.~Pincus, Europhys. Lett. {\bf 7}, 383 (1988).
% 
\bibitem{larson90} 
R.~G. Larson, E.~S.~Shaqfeh, and  S.~J.~Muller, 
J. Fluid Mech. {\bf 218}, 537 (1990); 
R.~G.~Larson, Rheol. Acta {\bf 31} , 213 (1992). 
% 
\bibitem{groisman96} 
A.~Groisman and V.~Steinberg, Phys. Rev. Lett. {\bf 77}, 1480 (1996); 
{\bf 78}, 1460 (1997); Europhys. Lett. {\bf 43}, 165 (1998).
% %
\bibitem{cerda97} 
E.~Cerda and E.~Tirapegui, Phys. Rev. Lett. {\bf 78}, 859 (1997); J. Fluid Mech.
{\bf 368}, 195 (1998). 
% 
\bibitem{muller97} 
H.~W.~M\"{u}ller, H.~Wittmer,  C.~Wagner,   J.~Albers, and  K.~Knorr,
Phys. Rev. Lett. {\bf 78}, 2357 (1997). 
% 
\bibitem{muller98a} H.~W.~M\"{u}ller and W.~Zimmermann, to appear in Europhys. Lett. (1998).
%  
%
\bibitem{edwards94} W.~S.~Edwards and S.~Fauve, J. Fluid Mech. , {\bf 278}, 
123 (1994).
%
\bibitem{chen97} 
P.~Chen and J.~Vinals, Phys. Rev. Lett. {\bf 79}, 2670 (1997). 
% 
\bibitem{kumar94} 
K.~Kumar and L.~S.~Tuckerman, J. Fluid Mech. {\bf 279}, 49 (1994). 
% 
\bibitem{muller93} 
H.~W.~M\"{u}ller, Phys. Rev. Lett. {\bf 71}, 3287 (1993).
%
% 
\end{thebibliography}
\end{document}